\documentclass[preprint,prd,aps,superscriptaddress,nofootinbib]{revtex4}

\usepackage{graphicx}

\begin{document}

\title {\large Thermal leptogenesis in a model with mass varying neutrinos}

\author{ Xiao-Jun Bi }
\affiliation{ Institute of High Energy Physics, Chinese Academy of
Sciences, P.O. Box 918-4, Beijing 100039, People's Republic of China}

\author{ Peihong Gu }
\affiliation{ Institute of High Energy Physics, Chinese Academy of
Sciences, P.O. Box 918-4, Beijing 100039, People's Republic of China}

\author{Xiulian Wang}
\affiliation{ Institute of theoretical Physics, Chinese Academy of
Sciences, P.O. Box 2735, Beijing 100080, People's Republic of China}

\author{ Xinmin Zhang }
\affiliation{ Institute of High Energy Physics, Chinese Academy of
Sciences, P.O. Box 918-4, Beijing 100039, People's Republic of China}

\date{\today}

\begin{abstract}

In this paper we consider the possibility of neutrino mass varying
during the evolution of the Universe and study its implications
on leptogenesis. Specifically, we take the minimal seesaw model of neutrino
masses and introduce a coupling between the right-handed neutrinos
and the dark energy scalar field, the Quintessence. 
In our model, the right-handed
neutrino masses change as the Quintessence scalar evolves. We then
examine in detail
the parameter space of this model allowed by the observed baryon number
asymmetry. Our results show that it is possible
to lower the reheating temperature in this scenario in comparison with
the case that the neutrino masses are unchanged, which
helps solve the gravitino problem. Furthermore, 
a degenerate neutrino mass patten with 
$m_i$ larger than the upper limit given in the minimal leptogenesis
scenario is permitted.

\end{abstract}

\maketitle

\section {introduction}

The baryon number asymmetry of the Universe has been determined
to the precision of less than 10\% after the first year observations of
the Wilkinson Microwave Anisotropy Probe (WMAP) experiment\cite{wmap}.
The reported value of the asymmetry is
\begin{equation}
\eta_B\equiv \frac{n_B}{n_\gamma}=6.1^{+0.3}_{-0.2}\times 10^{-10}\, ,
\end{equation}
where $n_B=n_b-n_{\bar{b}}$ and $n_\gamma$ are the baryon and photon
number densities, respectively.

On the theoretical side, many models have been proposed in the
literature\cite{rev} to explain the small while non-zero $\eta_B$
dynamically. Among various
models, leptogenesis is one of the most attractive
scenarios\cite{lepto}, where lepton number asymmetry is 
converted to baryon number asymmetry via the $(B+L)$-violating
sphaleron interactions\cite{spha}. This mechanism 
has been studied extensively in the literature\cite{early,vari,ibarra,
buch,new,therm}.

In the minimal scenario of leptogenesis, the standard model is
extended by including three generations of right-handed (RH)
neutrinos, which are Majorana fermions and couple to the
lepton doublets through
Yukawa couplings. The heavy RH neutrinos are
produced in the early universe by thermal scattering with the primordial
thermal bath via the Yukawa coupling.
In this framework, the cosmological baryon asymmetry
is naturally connected to the neutrino properties through
the seesaw mechanism\cite{seesaw}, which is the most natural
explanation for the tiny neutrino masses observed in the
neutrino oscillation experiments\cite{neu}. As shown in Ref. \cite{buch},
the amount of the baryon asymmetry generated in this minimal 
thermal leptogenesis scenario can be characterized by four
parameters: the maximal CP asymmetry $\epsilon_1$, the lightest RH neutrino
mass $M_1$, the effective light neutrino mass $\widetilde{m}$ and the
quadratic mean of the light neutrino masses $\overline{m}$,
under the assumption that the dominant contribution to the lepton
asymmetry is given by the decay of the lightest RH neutrino, $N_1$.
The detailed calculations show that
to produce the observed baryon asymmetry, it requires $M_1 \gtrsim 10^{10}
GeV$ and $\overline{m} \lesssim 0.2 eV$ which corresponds to
$m_i \lesssim 0.12 eV$.  

Although it is theoretically elegant and simple,
this scenario seems to need a high reheating temperature, $T_R \sim M_1
\gtrsim \mathcal{O}(10^{10} GeV)$, which is only marginally compatible
with the bound from the gravitino problem, $T_R \le (10^8 - 10^{10}) GeV$, 
in supersymmetric models\cite{grav}. 
The degenerate solution of the neutrino masses
is also strongly disfavored in this scenario. Note that if the evidence
of neutrinoless double $\beta$ decay with $m_{ee} = (0.05-0.86) eV$ \cite{0b}
(at 95\% C.L.) is confirmed, a degenerate neutrino spectrum is required.
A recent study on the cosmological data also showed a preference
for degenerate neutrinos with $m_i=0.2 eV$\cite{deg}.
A possible way to accommodate the degenerate neutrinos by the thermal
leptogenesis is the resonant enhancement
of the CP asymmetry when the heavy RH neutrinos are nearly degenerate,
$|M_{2,3}-M_1|=\mathcal{O}(\Gamma_i)$\cite{vari,new}.

In the standard $\Lambda$CDM cosmology, the dark energy which
drives the universe accelerating weights about $73\%$ of the total
energy of the universe\cite{wmap,sup,sdss}. However, the nature
of the dark energy (DE) remains mysterious. It could be simply a
remnant small cosmological constant. However, many physicists are
attracted by the dynamical solution with a scalar field (or multi-scalar
fields) like Quintessence. Being a dynamical component, the scalar
field dark energy is expected to interact with the ordinary
matter. In the literature there have been a lot of studies on the
possible couplings of Quintessence to baryon, dark matter and
photons\cite{int}. Recent data on the possible variation of the
electromagnetic fine structure constant reported in\cite{webb} has
triggered interests in studies related to the interactions between
Quintessence and the matter fields.

Recently we have studied various possible interactions,
with or without supersymmetry, between Quintessence and
the matter field of the electroweak standard model\cite{li3}. In a
previous paper\cite{wangxl} we considered a possibility of
neutrinos coupling with Quintessence and studied its implications on
cosmology. In this paper we propose a scenario that the right-handed neutrinos
in the minimal seesaw model  couple
to Quintessence and examine its implications on the thermal
leptogenesis. Other possible implications of varying neutrino
masses in astrophysics and cosmology are discussed in
Ref. \cite{nelson}.
With a detailed numerical calculation we will show that
in this scenario the reheating
temperature can be as low as about $10^8 GeV$ and the degenerate
neutrino mass patten with $m_i\sim \mathcal{O}(0.2 eV)$ will be allowed.

This paper is organized as follows. In Sec II, we will present the
framework and formulation for the calculation of 
leptogenesis in our model. We will explain how the interaction between the
right-handed neutrinos and Quintessence affect
$\eta_B$ in terms of the four parameters
$\epsilon_1$, $M_1$, $\widetilde{m}$ and $\overline{m}$.
In section III, we will present the numerical results.
In section IV, we consider a Quintessence model and propose
a specific example of Quintessence to RH neutrinos, then study
the interplay between the leptogenesis and the evolution of Quintessence.
Sec. V is the summary and conclusion of this paper.

\section{Interacting dark energy scalar field with neutrinos}

\subsection{The model}

We extend the minimal seesaw model by introducing a new coupling
between Quintessence scalar field, $Q$, and the RH neutrinos.
The Lagrangian relevant to leptogenesis is given by
\begin{equation}
\mathcal{L}=\mathcal{L}_{lep}+\mathcal{L}_Q\, ,
\end{equation}
with
\begin{equation}
-\mathcal{L}_{lep}= Y_{ij}\bar{L}_i\tilde{H}
N_j + \frac{1}{2}M_i(Q)N^T_iCN_i + h.c.\, ,
\end{equation}
and
\begin{equation}
\mathcal{L}_Q=\frac{1}{2}\partial_\mu Q\partial^\mu Q-V(Q)\, ,
\end{equation}
where $Y_{ij}$ is the Yukawa coupling of the RH neutrinos, $M_i(Q)$
is the Majorana masses of the RH neutrinos which is now a function
of the value of $Q$, and $V(Q)$ is the potential of the Quintessence.
We have taken the basis that the RH Majorana mass matrix is diagonal.
The form of $V(Q)$
for a specific model and an explicit form of $M(Q)$ will be given
in Sec. IV.  The
Quintessence dependent masses of the right-handed neutrinos give
rise to different Majorana masses at the early epoch 
from that at the present epoch, which is the key for the difference of
our model from the usual leptogenesis scenario.

\subsection{The CP asymmetry}

Generally, the Yukawa coupling $Y$ contains CP violating phases and
lead to different branching ratios for $N$ decays into
 lepton and antilepton.
The asymmetry for $N_1$ decays is\cite{early,vari}
\begin{eqnarray}
\epsilon_1 &=& \frac{\Gamma(N_1\to LH) - \Gamma(N_1\to
\bar{L}\bar{H})}{\Gamma(N_1\to LH) + \Gamma(N_1\to
\bar{L}\bar{H})} \nonumber \\
& \approx & -\frac{3}{16\pi}
\frac{1}{(Y^\dagger Y)_{11}}
\sum_{i=2,3} \text{Im}\left[ (Y^\dagger Y)_{1i} \right]^2\frac{M_1}{M_i}\ ,
 \;\; \text{for} \;\;
M_1\ll M_2, M_3 \; .
\end{eqnarray}

Since the light neutrino masses are generated by the seesaw
mechanism\cite{seesaw}
and constrained by the neutrino oscillation experiments, the CP
asymmetry is also constrained by the neutrino data. In the case of
hierarchical neutrinos, i.e., $m_3\approx \sqrt{\Delta m_{atm}^2}
\approx 0.05 eV
\gg \sqrt{\Delta m_{sol}^2} \sim 0.008 eV \gg m_1$, an approximate
upper bound on the CP asymmetry\cite{ibarra} is
\begin{equation}
|\epsilon_1| \lesssim \frac{3}{16\pi} \frac{M_1\sqrt{\Delta m_{atm}^2}}
{v^2} \ ,
\end{equation}
with $v=174 GeV$ being the vacuum expectation value of the neutral Higgs boson.
For the degenerate case, i.e., $m_1\approx m_2\approx m_3 \approx
\overline{m}/\sqrt{3} \gg
\sqrt{\Delta m_{atm}^2}$, the upper bound is given by
\begin{equation}
|\epsilon_1| \lesssim \frac{3\sqrt{3}}{16\pi} \frac{M_1\Delta m_{atm}^2}
{v^2\overline{m}} \ ,
\end{equation}
where $\overline{m}=\sqrt{m_1^2+m_2^2+m_3^2}$ is the quadratic mean of
the neutrino masses. One can see that the maximal $\epsilon_1$ is proportional
to $M_1$ while inversely proportional to $\overline{m}$. In the following
we always take the upper bound of $\epsilon_1$ to study
the maximal value of $\eta_B$ on the parameter space.

The final lepton asymmetry is determined in general by
\begin{equation}
\label{yl}
Y_L\equiv \frac{n_L-n_{\bar{L}}}{s} = \kappa \frac{\epsilon_1}{g_*}\ ,
\end{equation}
where $s$ is the entropy density, $g_*$ represents the number
of the relativistic
degrees of freedom at the time when $N_1$ decays and $\kappa < 1$ represents
the washout effects for the lepton number asymmetry in the thermal bath.
The lepton asymmetry in Eq. (\ref{yl}) will be converted partly 
into baryon asymmetry by the electroweak sphaleron process 
taking into account
the gauge, Yukawa interaction and the QCD sphaleron effects\cite{qcdspha},
$Y_B = a Y_{B-L}\approx -a Y_L$, with $a=28/79$\cite{therm,chem}.

\subsection{The washout effect}

The factor $\kappa$ in Eq. (\ref{yl}) can be calculated by solving 
the Boltzmann
equations\cite{buch,therm}. At the leading order of $\epsilon_1$ one
has the Boltzmann equations for $Y_{N_1}=n_{N_1}/s$ and
$Y_{B-L}=n_{B-L}/s$,
\begin{eqnarray}
\label{boln1}
\frac{sH}{z}\frac{dY_{N_1}}{dz} &=& -\left( \frac{Y_{N_1}}{Y_{N_1}^{eq}}
-1\right)\left( \gamma_D+2\gamma_{H,s}+4\gamma_{H,t} \right)\, ,\\
\label{bolbl}
\frac{sH}{z}\frac{dY_{B-L}}{dz} &=& -\epsilon_1 \gamma_D
\left( \frac{Y_{N_1}}{Y_{N_1}^{eq}} -1\right) - \frac{Y_{B-L}}{Y_L^{eq}}
\gamma_W\, ,
\end{eqnarray}
with
\begin{equation}
\gamma_W=\frac{1}{2}\gamma_D
+2\gamma_N+2\gamma_{N,t}+2\gamma_{H,t}+
\frac{Y_{N_1}}{Y_{N_1}^{eq}}\gamma_{H,s}\, ,
\end{equation}
where $\gamma$ is the space time density of the scatterings in thermal
equilibrium. In the equations above, $\gamma_D$ refers to
the reaction density for the process of $N_1$ decays (and inverse decays),
$N_1\leftrightarrow LH (\bar{L}\bar{H})$, $\gamma_{H,s}$ and
$\gamma_{H,t}$ for the processes of 
$\Delta L=1 $ scatterings via exchanging Higgs boson,
$N_1L(\bar{L})\leftrightarrow \bar{t}(t)Q(\bar{Q})$ in $s$ channel
and $N_1t(\bar{t}) \leftrightarrow \bar{L}(L)Q(\bar{Q})$ in $t$ channel,
$\gamma_N$ and $\gamma_{N,t}$ for the processes of 
$\Delta L=2 $ scatterings via exchanging the RH neutrinos,
$LH\leftrightarrow \bar{L}\bar{H}$ in both $s$ and $t$ channels
and $LL\leftrightarrow \bar{H}\bar{H}$, $\bar{L}\bar{L}\leftrightarrow HH$
in $t$ channel.

In Eq. (\ref{boln1}), the minus sign on the right-hand side means that
all the reactions drive the $Y_{N_1}$ to its equilibrium value
$Y_{N_1}^{eq}$.
In Eq. (\ref{bolbl}), the first term on the right-hand side is the source
term, which gives rise to non-zero $Y_{B-L}$, while the second term
represents washout effects that lead $Y_{B-L}$ to zero.

The $\gamma_N$ consists two parts,
the resonant part, $\gamma_{N,res}$ and the 
non-resonant part, $\gamma_{N,non}$,
i.e., $\gamma_N=\gamma_{N,res}+\gamma_{N,non}$. The
$\gamma_D$, $\gamma_{H,s}$, $\gamma_{H,t}$ and $\gamma_{N,res}$ are
all proportional to the effective neutrino mass\cite{buch}
\begin{equation}
\widetilde{m}=\frac{(Y^\dagger Y)_{11} v^2}{M_1}\ ,
\end{equation}
while $\gamma_{N,non}$ and $\gamma_{N,t}$ are proportional to
$M_1\overline{m}^2$. These properties are essential in understanding
the behavior of the solutions of the Boltzmann equations.
When $\delta \gamma_W
=\gamma_{N,non}+\gamma_{N,t}\propto M_1\overline{m}^2$ is negligible, 
i.e., at $M_1(\overline{m}/0.1 eV)^2
< 10^{13} GeV$\cite{buch}, the washout factor $\kappa$
is mainly determined by $\widetilde{m}$ and nearly
independent of $M_1$ and $\overline{m}$.
The effect of $\widetilde{m}$ has two different aspects. When it is very small
($\lesssim 10^{-3} eV$), the RH neutrinos will not be able to reach thermal
equilibrium via the scattering with the thermal bath, so $\kappa$ will be
 small.
Conversely, if $\widetilde{m}$ is too large the RH neutrinos can
be brought into thermal equilibrium rapidly. However, the washout effect
will also be large, consequently we have a small $\kappa$. 
Around $\widetilde{m}\sim
10^{-3} eV$, $\kappa$ reaches its maximal value.
When $\delta \gamma_W$
is sizable $\kappa$ will be suppressed exponentially,
since the washout effect is given by
\begin{equation}
\kappa (z) = \int_{z_{in}}^z dz' \gamma_D(z') \Delta (z')
e^{-\int_{z'}^z dz'' z''\gamma_W(z'')/(sHY_L^{eq})}\, ,
\end{equation}
with $\Delta(z)= (Y_{N_1}/Y_{N_1}^{eq}-1) z/(sH)$.
In this case, $\kappa$ becomes nearly independent of $\widetilde{m}$.

In short, $\epsilon_1$ (as well as $Y_L$) increases with $M_1$
while decreases with $\overline{m}$ linearly when
$\delta \gamma_W$ is negligible, $M_1(\overline{m}/0.1 eV)^2
< 10^{13} GeV$\cite{buch}. However, when $\delta \gamma_W\propto
M_1\overline{m}^2$ is sizable, $M_1(\overline{m}/0.1 eV)^2
> 10^{13} GeV$, the final asymmetry $Y_L$
will be suppressed exponentially\cite{buch}. 
The $\kappa$ reaches maximal at
$\widetilde{m}\sim 10^{-3} eV$ when the effect of $\delta \gamma_W$
is negligible. It is shown in Ref. \cite{buch} that
to generate the required value of the baryon number asymmetry, 
$\eta_B$,
one needs $M_1\gtrsim 10^{10} GeV$ and $\overline{m}\lesssim 0.2 eV$ 
in general.

\subsection{The effect of Quintessence}

With the interaction between the Quintessence and the RH
neutrinos, $M_1(Q)$ is now a function of the Quintessence field.
$M_1$ at the epoch when it decays will differ from that at the
present epoch. We introduce a parameter $K$ 
\begin{equation}
K(Q^D)=\frac{M_1^0}{M_1^D},
\end{equation}
where $M_1^0=M_1(Q^0)$ and $M_1^D=M_1(Q^D)$ denote the Majorana
masses of the RH neutrinos at present and at the decay time. For
simplicity we take $K$, as well as $M_1$, as a constant 
when solving the Boltzmann equations. In section IV, we will
show that $Q$ is quite flat during the period of leptogenesis 
for the model we consider  and this is a 
reasonable assumption. We then have
\begin{equation}
m_\nu^D=-Y\frac{1}{M^D}Y^Tv^2=m_\nu^0\cdot K\ ,\;\;
\overline{m}^D=\overline{m}^0\cdot K\ ,\;\;
\widetilde{m}^D=\widetilde{m}^0\cdot K\ ,
\end{equation}
where the masses with the superscripts ``D" and ``0" correspond to
the values evaluated at the leptogenesis and at the present time
respectively.

\section{numerical results}

In this section we present our numerical results by solving the Boltzmann
equations. Besides $M_1$, $\overline{m}$ and $\widetilde{m}$ we now
introduce a new parameter $K$. We will show that varying $K$ 
makes it possible for
$M_1$ to be as low as about $10^8 GeV$ and $\overline{m}$ can be
as large as $0.4 eV$ which corresponds to 
$m_i \approx 0.23 eV$\cite{wmap}\footnote{
This is the upper bound set by WMAP.
Recent SDSS\cite{sdss} lowers this limit to $0.2 eV$. When neutrino
masses vary during the evolution of the Universe, however, these cosmological
bounds on the absolute neutrino masses may be different which is 
worthwhile studying further.}.


\begin{figure}
\includegraphics[scale=0.4]{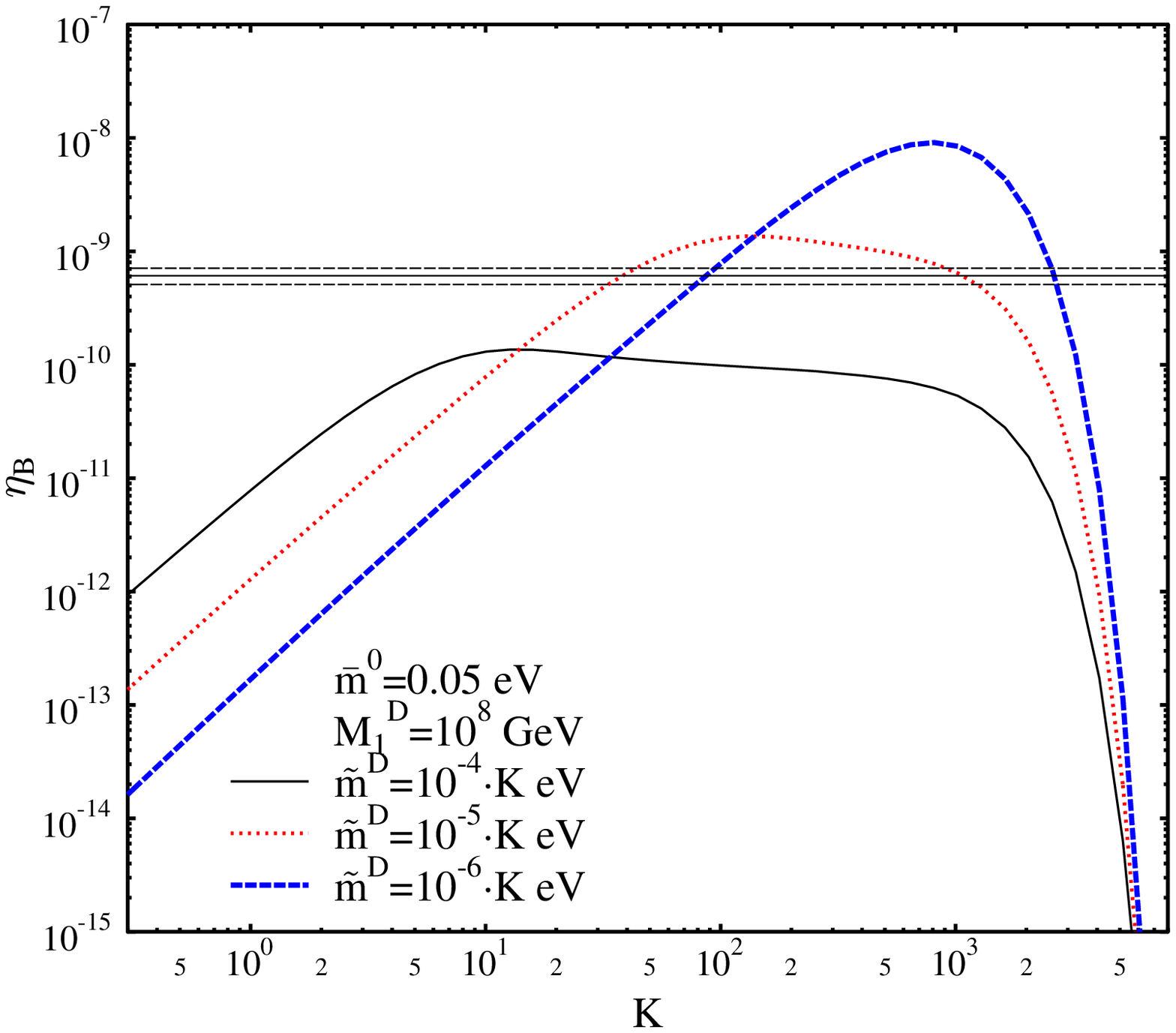}
\includegraphics[scale=0.4]{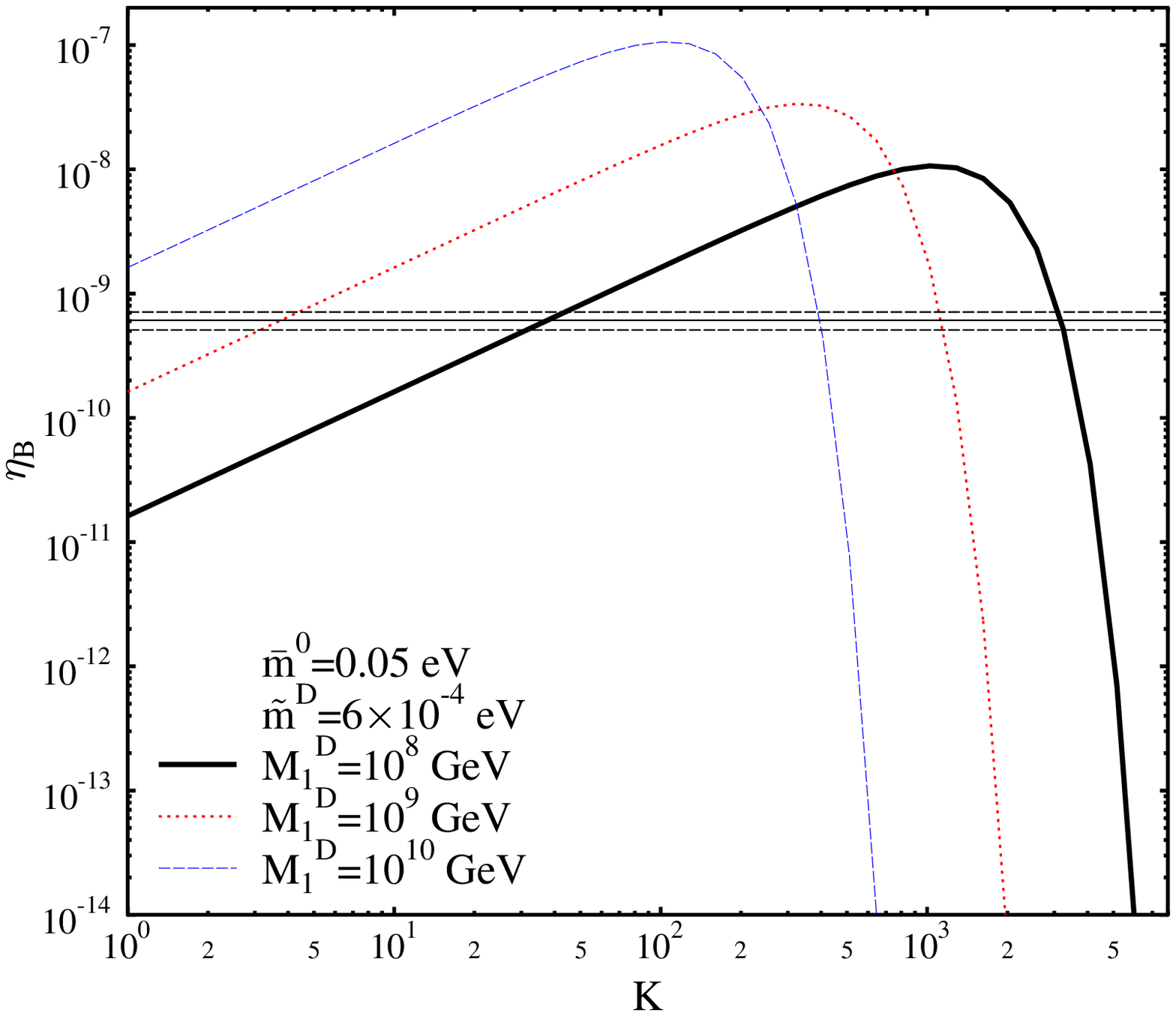}
\caption{\label{fig1}
The baryon number asymmetry $\eta_B$ as a function of $K$ for the 
hierarchical neutrino mass spectrum, i.e., $\overline{m}^0=0.05 eV$.
On the left panel, we take $M_1^D=10^8 GeV$ and
$\widetilde{m}^0=10^{-4}, 10^{-5}, 10^{-6} eV$ respectively.
On the right panel, we take $\widetilde{m}^D=6\times 10^{-4} eV$
and $M_1^D=10^8, 10^9, 10^{10} GeV$ respectively. The horizontal
lines represent $\eta_B=5.1, 6.1, 7.1 \times 10^{-10}$ respectively.
}
\end{figure}

In Fig. \ref{fig1}, we plot the baryon number asymmetry $\eta_B$ as a function
of $K$ for the hierarchical neutrino mass spectrum, i.e.,
$\overline{m}^0=0.05 eV$. On the left panel, we take $M_1^D=10^8 GeV$
for $\widetilde{m}^0=10^{-4}, 10^{-5}, 10^{-6} eV$ respectively.
One can see from the figure that 
$\eta_B$ increases with $K$ linearly until $\widetilde{m}^D \approx
10^{-3} eV$, i.e., $\eta_B$ increases with
$K$ until $K\approx 10^1, 10^2, 10^3$ for
$\widetilde{m}^0=10^{-4}, 10^{-5}, 10^{-6} eV$ respectively. In these regions,
the amount of enhancement of $\epsilon_1$ as $K$ gets large
dominates over that of
the washout effect so that $\eta_B$ increases linearly with $K$.
However, as the $\overline{m}^D$ is about $\overline{m}^0\cdot K
\sim  0.05\cdot 10^3 eV$, $\delta \gamma_W$ becomes sizable then
$\eta_B$ decreases exponentially as $K$. In the region between these
two points, $\eta_B$ is a flat function of $K$. This is
due to the effect of $K$ which enhances $\epsilon_1$ on one hand
while decrease $\kappa$ as $\widetilde{m}^D\gtrsim 10^{-3} eV$
on the other hand.
On the right panel, we fix $\widetilde{m}^D=6\times 10^{-4} eV$, which
gives the maximal $\kappa$ for the negligible $\delta \gamma_W$\cite{buch}, 
and take 
$M_1^D= 10^8, 10^9, 10^{10} GeV$
respectively. One can see from this figure that 
$\eta_B$ increases with $K$ linearly until
$M_1^D (\overline{m}^D/0.1 eV)^2 \sim 10^{13} GeV$.
The different behavior between the two figures is easy to understand
if one notices that $\widetilde{m}^D$ is fixed and does not increase
with $K$ on the right panel.

\begin{figure}
\includegraphics[scale=0.5]{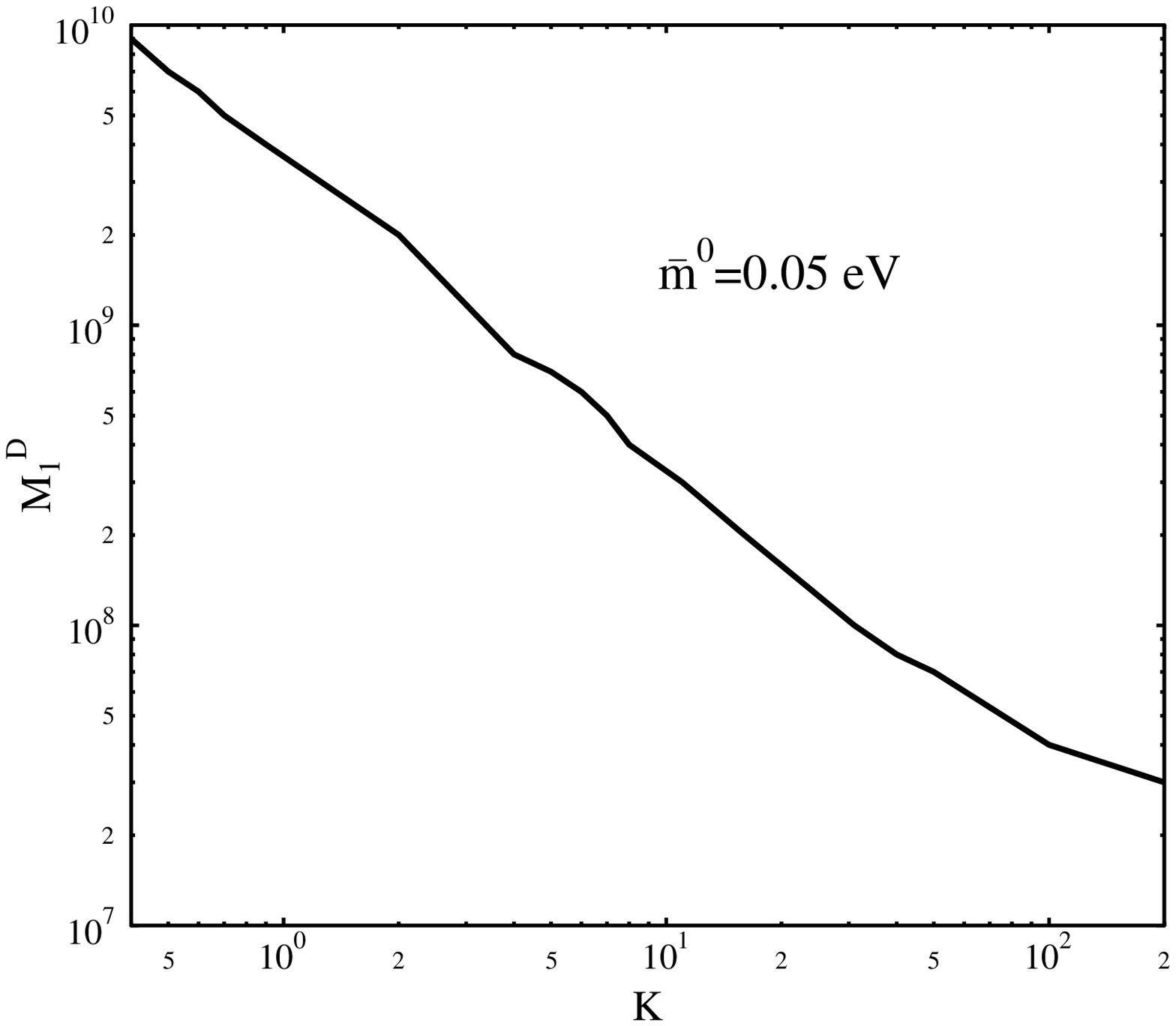}
\caption{\label{fig2}
The contour of $\eta_B=5.5\times 10^{-10}$ in the $M_1^D - K$ plane
for the hierarchical neutrino spectrum, $\overline{m}^0=0.05 eV$.
}
\end{figure}

In Fig. \ref{fig2}, we plot the contour for $\eta_B=5.5\times
10^{-10}$ in the $M_1^D - K$ plane for the hierarchical neutrinos.
One can see from this figure that as K gets large, $M_1^D$ is allowed to take
a smaller value.
For example, at $K\approx 30$, $M_1^D$ can be as low as about $10^8 GeV$.
As a result of it, the reheating temperature required will be 
lowered than that in the minimal thermal leptogenesis.

\begin{figure}
\includegraphics[scale=0.5]{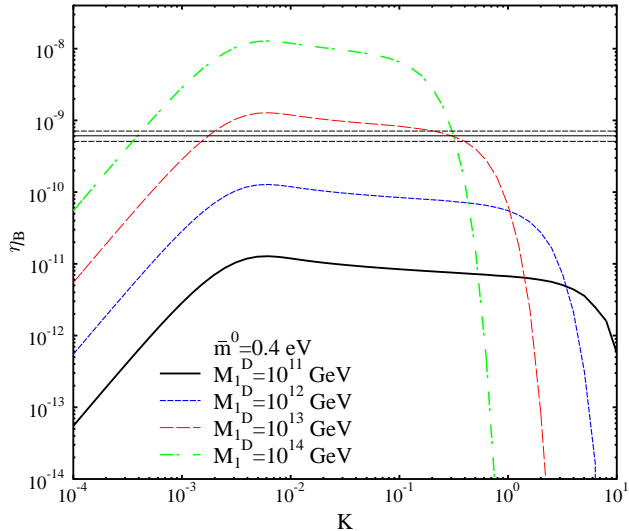}
\caption{\label{fig3}
The baryon number asymmetry $\eta_B$ as a function of $K$ for the degenerate
neutrino mass spectrum, $\overline{m}^0=0.4 eV$.
We take $M_1^D=10^{11}, 10^{12}, 10^{13}, 10^{14} GeV$
respectively.  The horizontal
lines represent $\eta_B=5.1, 6.1, 7.1 \times 10^{-10}$ respectively.
}
\end{figure}

In Fig. \ref{fig3}, we plot $\eta_B$ as a function of $K$ for the degenerate
neutrinos, taking $\overline{m}^0=0.4 eV$ ($m_i\approx 0.23 eV$) for an example.
It should be noticed that $\widetilde{m}$ is constrained as
$m_1 <\widetilde{m} < m_3$ if no fine tunning is taken\cite{buch}.
We take $\widetilde{m}^0\approx m_i^0\approx 0.23 eV$ in this case.
The figure looks similar to that for the hierarchical case.
For $\widetilde{m}^D =(0.23 \cdot K) eV\lesssim 10^{-3} eV$,
$\eta_B$ increases linearly with $K$. At the point
$M_1^D (\overline{m}^D/0.1 eV)^2 \sim 10^{13} GeV$, $\eta_B$ starts to
decrease exponentially as $K$. In the region between, $\eta_B$ is almost
a flat function of $K$. We notice that for $K=1$, which corresponds
to the case of vanishing Quintessence coupling, no parameter space will
satisfy the observed $\eta_B$. However, for $M_1^D=10^{13} GeV$, $\eta_B$
can meet the requirement for $K\lesssim 0.5$, which is not far from
$K=1$.
One can see that non-zero $K$ makes it possible to generate $\eta_B$ required
for the degenerate neutrino mass spectrum. However, in this case
we have to take $K$ to be smaller than $1$ so that $\kappa$ is not
suppressed exponentially. This leads to a small $\epsilon_1$ too, which
forces us to take a bigger $M_1^D$. Therefore, it will not be possible to lower
$M_1^D$ and increase $\overline{m}^0$ simultaneously.

\begin{figure}
\includegraphics[scale=0.5]{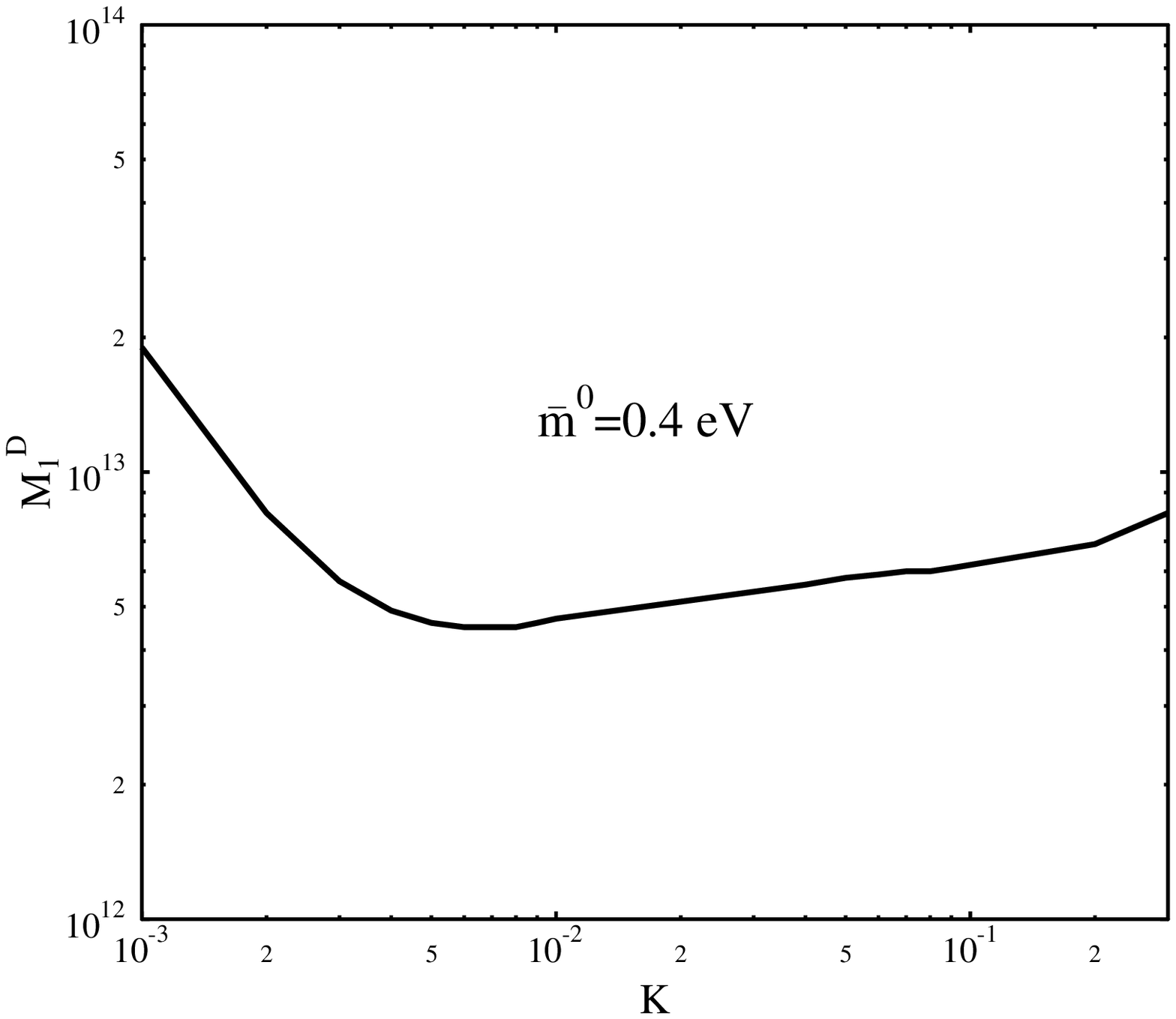}
\caption{\label{fig4}
The contour of $\eta_B=5.5\times 10^{-10}$ in the $M_1^D - K$ plane
for the degenerate neutrino spectrum, $\overline{m}^0=0.4 eV$.
}
\end{figure}

In Fig. \ref{fig4} we plot the contour of $\eta_B=5.5\times 10^{-10}$ 
in the $M_1^D - K$ plane
for the degenerate neutrinos. One can see that
$M_1^D$ decreases linearly firstly, then starts to increases
slowly. At $K\approx 0.007$, $M_1^D$ reaches its minimal value, 
$M_1^D\approx 4.5 \times 10^{12} $GeV.

\section{Quintessence models}

In the last section we present our numerical results of leptogenesis
taking $K$ as a free parameter. In this section we will consider 
a specific Quintessence model
and propose a specific form of its coupling to the RH neutrinos.
We will study numerically the evolution of $Q$ taking into account
the back reaction of the neutrino background and calculate the factor
$K$.

In the literature, there have been various proposals for the explicit
form of couplings in studying the interaction between the Quintessence
and the matter field. For example, one usually introduces
$QF_{\mu\nu}F^{\mu\nu}$ ($F_{\mu\nu}$ is the electromagnetic field
strength tensor) to study the variation of the
electromagnetic fine structure constant. However, in an attempt to
understand the puzzle why the density of dark matter and dark
energy are nearly equal today, the authors of Ref. \cite{dmint} recently
consider a model of interacting Quintessence with dark matter
and in their scenario the mass of dark matter particle depends 
exponentially on the value of $Q$, $m(Q)=\bar{m} e^{-\lambda Q/M_{pl}}$.

In this paper, we assume that the coupling between Quintessence 
and the RH neutrinos 
takes a simple form as
\begin{equation}
\label{mq}
M_i(Q)=\overline{M}_i e^{\beta \frac{Q}{M_{pl}}} ,
\end{equation}
where $\beta$ is a $\mathcal{O}(1)$ coefficient. 
The ratio $K$ is then given as
\begin{equation}
K\equiv \frac{M_i^0}{M_i^D}= e^{\beta \frac{Q^0
-Q^D}{M_{pl}}}\ .
\end{equation}

\begin{figure}
\includegraphics[scale=.5]{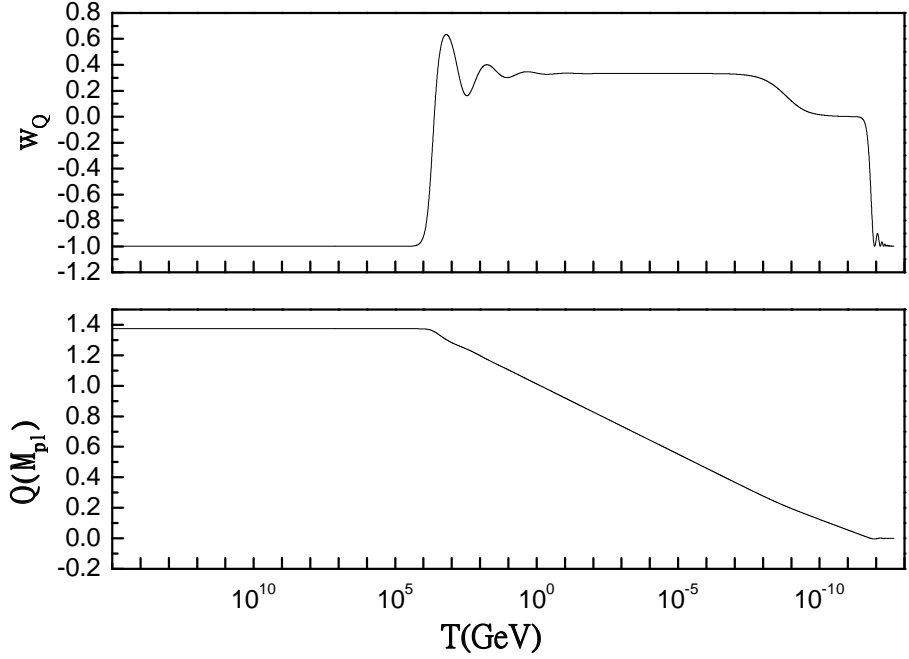}
\caption{\label{2exp}
The evolution of $W_Q$ and $Q$ as a function of 
the temperature $T$ for the double
exponential Quintessence model without the coupling with the RH neutrinos. }
\end{figure}

For a numerical calculation of $K$, we consider a Quintessence model 
with the double exponential potential\cite{Barreiro}
\begin{equation}
V=V_0(e^{\lambda Q}+e^{\alpha Q})\ .
\end{equation}
This model has the tracking property for suitable parameters.
In the absence of coupling in Eq. (\ref{mq}),
the evolution of Quintessence is described by the equations
\begin{eqnarray}
&&H^2 =\frac{8 \pi G}{3} (\rho_m +\rho_\gamma +\rho_Q),\;\\
&&\ddot{Q}+3H\dot{Q}+\frac{dV(Q)}{dQ}=0\ , 
\end{eqnarray} 
where $H$ is the Hubble constant,
$\rho_m$, $\rho_\gamma$ and $\rho_Q$ represent the energy densities of
matter, radiation and Quintessence respectively. We choose the
model parameters as $\lambda =100M_{pl}^{-1}$, $\alpha
=-100M_{pl}^{-1}$, the initial value of Quintessence field
$Q_i=1.374 M_{pl}$ and for the state of equation,
which is defined as 
\begin{equation}
\label{wq}
W_Q=\frac{p_Q}{\rho_Q}=\frac{{\dot{Q}^2}/{2}-V(Q)}
{{\dot{Q}^2}/{2}+V(Q)}\ \ ,
\end{equation}
the initial value is $W_{Q_i}=-1$.
We obtain that $\Omega_{Q_0} \simeq 0.72$ and 
the present equation of state of Quintessence is $W_{Q_0}\approx -1$ which
are consistent with the observational data.
In Fig. {\ref{2exp}} we show the evolution of $W_Q$ and $Q$ with 
the temperature $T$. One can see that Quintessence begins to
track the background at $T \lesssim 10^4 GeV$.


\begin{figure}
\includegraphics[scale=.5]{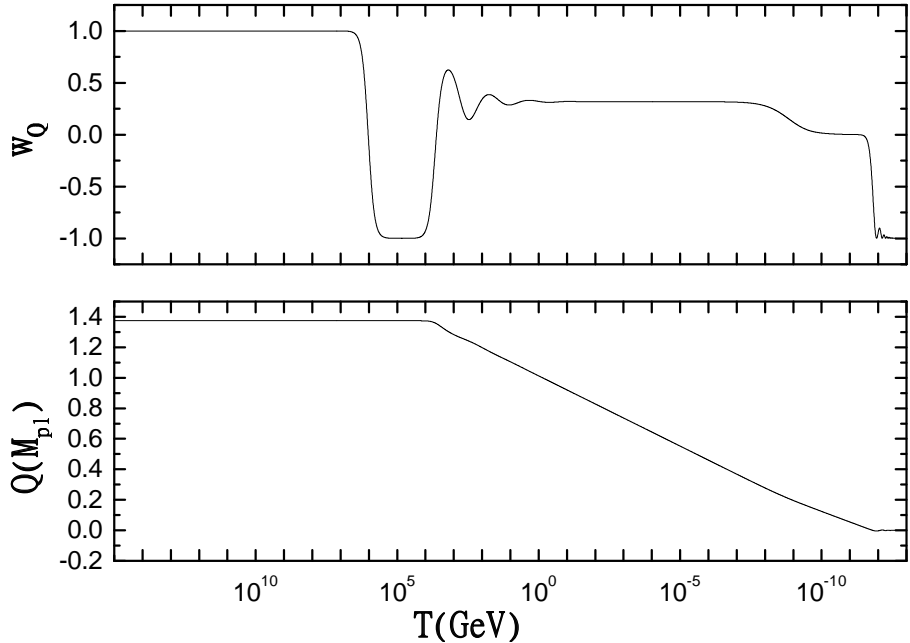}
\caption{\label{2exp31}
The evolution of $W_Q$ and $Q$ as a function of 
the temperature $T$ for the double
exponential Quintessence model including the coupling with the RH neutrinos. 
We take $\beta=-2.5$ and $\overline{M}_1=3.1 \times 10^9 GeV$.
We take the same definition of $W_Q$ as in Eq. (\ref{wq}).}
\end{figure}

\begin{figure}
\includegraphics[scale=.5]{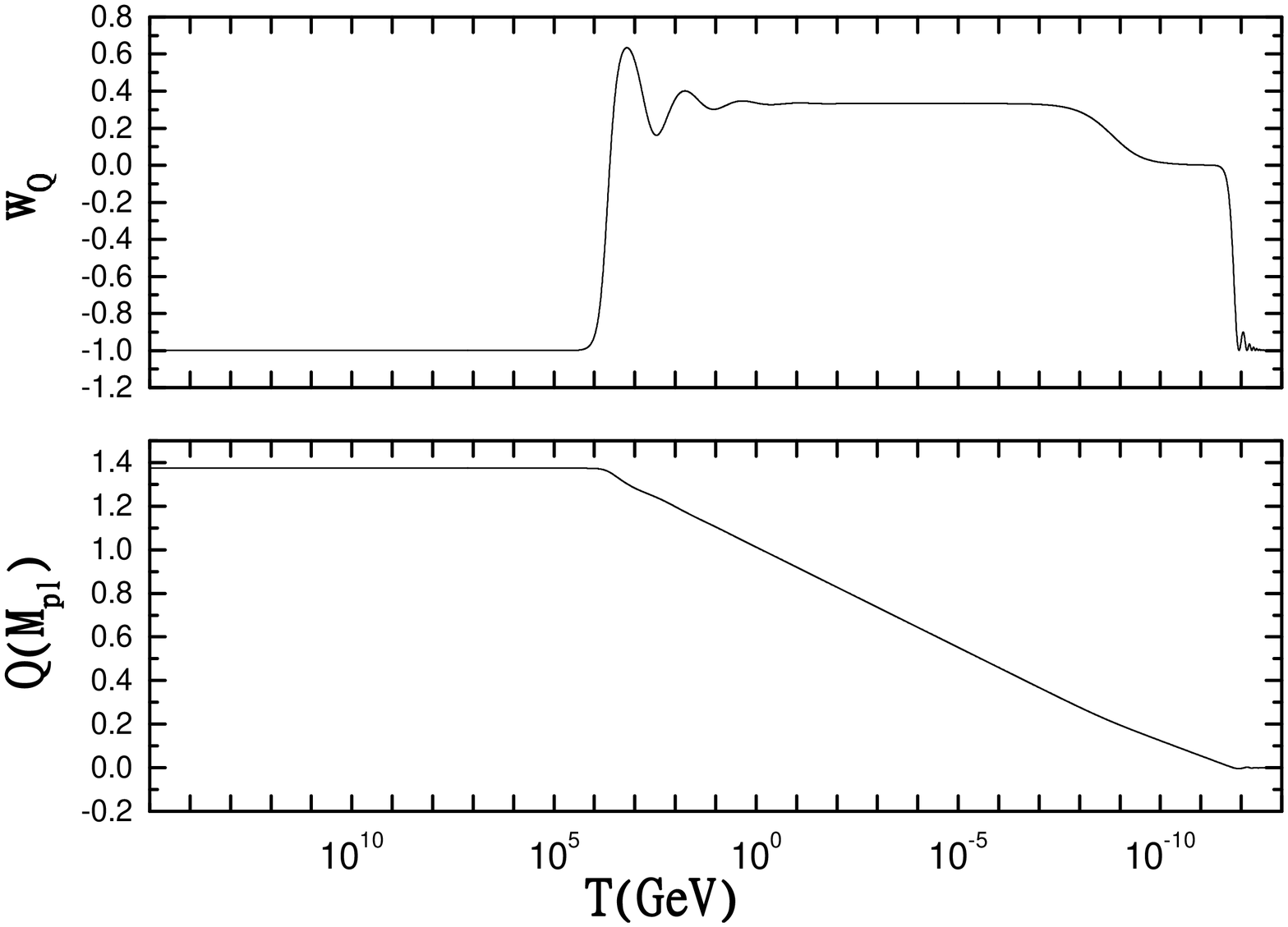}
\caption{\label{2exp007}
The evolution of $W_Q$ and $Q$ as a function of 
the temperature $T$ for the double
exponential Quintessence model including the coupling with the RH neutrinos. 
We take $\beta=3.62$ and $\overline{M}_1=3.2 \times 10^{10} GeV$.
We take the same definition of $W_Q$ as in Eq. (\ref{wq}).}
\end{figure}

Taking into account the interaction with the RH neutrinos,
we get the equation of evolution of Quintessence as
\begin{equation}
\label{vi}
\ddot{Q}+3H\dot{Q}+\frac{dV(Q)}{dQ}+\frac{dV_I(Q)}{dQ}=0\ .
\end{equation} 
The source term in the equation above is given by\cite{peebles}
\begin{equation}
\frac{dV_I(Q)}{dQ}=\sum_in_i\frac{dM_i}{dQ}
\left\langle\frac{M_i}{E}\right\rangle
=\frac{\beta}{M_{pl}}\frac{1}{\pi^2}T\sum_i M_i^3K_1(M_i/T)\ ,
\end{equation}
where $n_i$ and $E$ are the number density and energy of the RH neutrinos 
respectively, $\langle \rangle$ indicates
thermal average. In the last step of the equation we have taken the 
Maxwell-Boltzmann distribution of the RH neutrinos for
simplicity and $K_1$ is the modified Bessel function.
The energy density of Quintessence is taken the same form as
$\rho_Q=\dot{Q}^2/2+V(Q)$ in this case.

We then solve the equation (\ref{vi}) numerically, 
assuming $\overline{M}_3=10\overline{M}_2=100\overline{M}_1$.
The numerical results are shown in Figs. \ref{2exp31} and \ref{2exp007}.
In Fig. \ref{2exp31}, we take 
$\overline{M}_1 =3.1 \times 10^9$ GeV, $\beta=-2.5$, which
gives rise  to $Q_0 \approx 0$ and $Q_D =1.374 M_{pl}$. 
We then have $K\approx 31$ and $M_1^D \approx 10^8 GeV$, corresponding to
the case we considered in the hierarchical neutrino spectrum.
In Fig. \ref{2exp007}
we choose the parameters $\overline{M}_1 =3.15 \times 10^{10} GeV$
 and $\beta=3.62$. We find 
the values of $Q_0$ and $Q_D$ are almost the same as the above case.
We then have $K\approx 0.007$
and $M_1^D \approx 4.5\times 10^{12} GeV$, corresponding to the case 
that satisfy our requirement for the degenerate neutrinos. 

Comparing Figs. \ref{2exp31} and \ref{2exp007} with Fig. \ref{2exp},
one can see that the interaction of Quintessence with the RH neutrinos
does change the early behavior of the Quintessence field $Q$ and its
equation of state, however, does not change the tracking properties of this
model. Furthermore, the value of Quintessence field $Q$ remains
a constant in this model 
until $T\sim 10^4 GeV$ which satisfies our assumption for
a constant $K$ during the period of leptogenesis.

\section{Summary and conclusion}

The explanation of the baryon number asymmetry of the Universe remains
a challenge for cosmological and particle physics. Among various
theoretical models, leptogenesis is one of the most attractive scenarios.
Especially, the minimal thermal leptogenesis only needs the minimal
extension of the standard model, while the extension seems necessary
in order to explain the neutrino oscillation experiments.

The minimal thermal leptogenesis is closely connected with the properties
of the light neutrinos. Detailed studies show that, to generate the observed
baryon asymmetry, the constraints
$M_1 \gtrsim 10^{10} GeV$ and $m_i \lesssim 0.1 eV$ should be satisfied.
These results disfavor the degenerate neutrino mass spectrum strongly and
require a high reheating temperature.

We investigate in this paper 
the implications on the minimal thermal leptogenesis of
a new scenario that the RH neutrinos couple
with the Quintessence, whose evolution causes the masses of the RH
neutrinos vary. We study the interplay of this coupling between
the evolution of Quintessence and the leptogenesis. 
By solving the Boltzmann equations numerically
we show that $M_1$, at the decay time, can be as
low as about $10^8 GeV$ and $m_i$, at the present epoch, can
be as large as $m_i\sim 0.2 eV$.

\begin{acknowledgments}
We thank B. Feng for helpful discussions about the Quintessence
evolution. This work is supported in part by the National Natural
Science Foundation of China under the Grand No. 10105004,
19925523, 10047004 and the Ministry of Science and Technology of
China under Grant NO. NKBRSF G19990754. X. J. Bi is also supported
in part by the China Postdoctoral Science Foundation.
\end{acknowledgments}

\end{document}